# A Leap-on-Success Exhaustive Search Method to Find Optimal Robust Minimum Redundancy Arrays (RMRAs): New Array Configurations for Sensor Counts 11 to 20

Pradyumna Kunchala[1] and Ashish Patwari[2*]

[1]School of Computer Science and Engineering, Vellore Institute of Technology, Vellore, Tamil Nadu, 632 014, India

[2]School of Electronics Engineering, Vellore Institute of Technology, Vellore, Tamil Nadu, 632 014, India

*Corresponding Author Email ID: ashishpvit@gmail.com; ashish.p@vit.ac.in

**Abstract:** Two-fold redundant sparse arrays (TFRAs) are designed to maintain accurate direction estimation even in the event of a single sensor failure, leveraging the deliberate coarray redundancy infused into their design. Robust Minimum Redundancy Arrays (RMRAs), a specialized class of TFRAs, optimize this redundancy to achieve the maximum possible aperture for a given number of sensors. However, finding optimal RMRA configurations is an NP-hard problem, with prior research reporting optimal solutions only for arrays of up to ten sensors. This paper presents newly discovered optimal RMRA configurations for array sizes 11 to 15, identified using a novel Leap-on-Success exhaustive search algorithm that efficiently reduces computational effort by terminating the search upon locating optimal solutions. The robustness of these arrays was validated under all single-element failure scenarios using MATLAB simulations, confirming their superior resilience compared to some existing TFRAs vulnerable to failures at specific sensor positions. Furthermore, near-optimal configurations for array sizes 16 to 20 are also reported, highlighting the potential applicability of the proposed method for larger array designs given sufficient computational resources. This work not only advances the state-of-the-art in RMRA design but also introduces an effective search methodology that can be leveraged for future explorations in array configuration optimization.

## 1. Introduction

Sensor arrays are widely used in radar, sonar, seismology, acoustics, medical imaging, and wireless communications to estimate the direction of incoming wavefields [1], [2], [3]. For a fixed sensor count, sparse arrays outshine uniform/full arrays in terms of array aperture, degrees of freedom (DOF), angular resolution, and immunity to mutual coupling [4], [5], [6].

The advent of coprime and nested arrays in the year 2010 led to remarkable progress in the field of sparse array design [7], [8]. However, most array designs focused on maximizing the hole-free span of the difference coarray (DCA) or improving the array's immunity against mutual coupling or both [9], [10], [11], [12], [13], [14]. An important factor i.e., robustness to sensor failures, had been largely ignored in many designs [15].

Two-fold redundant arrays (TFRAs) contain physical sensors at positions that render a doubly redundant DCA [16]. Redundancy is carefully woven into the design of these arrays such that the array contains at least two sensor pairs to generate a given lag, that is, each spatial lag (difference) in the DCA is generated at least twice. This lag multiplicity ensures that the failure of a single sensor does not prevent any lag from occurring as it can obtained from alternate/redundant paths. In short, a TFRA shall provide - (i) a doubly redundant DCA during the healthy case and (ii) a hole-free DCA in the event of any single sensor failure, as illustrated in Fig. 1.

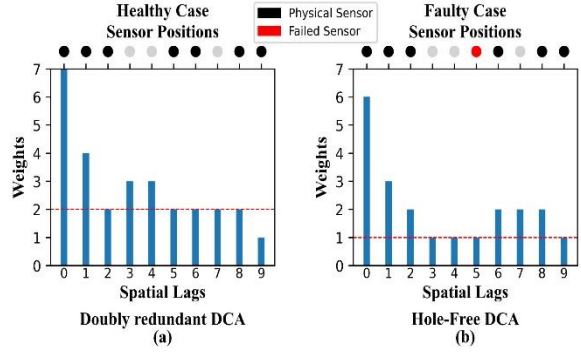

Fig. 1. Properties of a TFRA

Liu and Vaidyanathan proposed a variant of TFRAs, namely, the robust minimum redundancy array (RMRA) that offers the largest aperture ($L$) for a given number of sensors ($N$), while satisfying the conditions in Fig. 1. The RMRA lacks closed-form expressions (CFEs) for optimum sensor positions and has to be found using exhaustive search methods [17]. RMRA configurations for $N \leq 10$ were found via integer programming. However, the computational burden was prohibitive for finding larger arrays. As an alternative to RMRAs, Liu and Vaidyanathan proposed sub-optimal solutions such as the symmetric nested array (symNA) and the composite Singer array (CSA) that had CFEs for sensor positions [17], [18]. However, neither of these arrays can be defined for all array sizes, thereby limiting their practical applicability.

Around the same time, Zhu *et. al* proposed 2-fold redundant arrays (2FRAs) with similar goals. Although the fireworks algorithm and particle swarm optimization were initially used to determine 2FRAs, Zhu *et. al* soon proposed the idea of using double difference bases (DDBs) available in number theory as 2FRAs [16], [19]. However, it was later found that some DDB-converted-2FRAs are vulnerable to single sensor failures at specific positions, thus breaching the robustness requirements [20]. The DOA estimation performance of such arrays depends on the position of the failed sensor and is hence uncertain. Hence, a double difference alone cannot guarantee the intended two-fold redundancy and robustness against single sensor failures. We reinforce this with example array configurations in section 4.

TFRA refers to the entire family of sparse arrays that provide doubly redundant DCA in the healthy case and a hole-free DCA during single sensor failures. 2FRA refers to the version proposed by Zhu et al. Table 1 summarizes the current state-of-the-art in TFRA literature. As seen, there is no one-size-fits-all solution among the existing arrays, and hence, the field is still open for research.

Table 1

Comparison of Existing TFRAs

|  | SymNA | CSA | 2FRA | RMRA |
| --- | --- | --- | --- | --- |
| Defined for all array sizes | No | No | Yes | Not Applicable |
| CFEs for Sensor Positions | Yes | Yes | Yes | No |
| Truly Robust to all single-sensor failures | Yes | Yes | No | Yes |

Given the drawbacks of existing TFRA designs and the need for fault-tolerant DOA estimation in future applications [21], there is a need to find new RMRAs such that array designers have a large catalogue at their disposal. Hence, we attempted to solve the combinatorial optimization problem of finding optimal RMRAs using an exhaustive search procedure. Although exhaustive search methods, meta heuristic algorithms, and integer linear programming have been widely explored for optimizing minimum redundancy arrays (MRAs) and non-redundant arrays (NRAs) [22], [23], [24], [25], [26], [27], [28], the field of RMRA optimization is still in its infancy. To date, no systematic methods have been developed to solve the combinatorial problem of identifying optimal Robust Minimum Redundancy Arrays (RMRAs). This work addresses this gap by introducing a novel Leap-on-Success exhaustive search algorithm capable of discovering optimal configurations. To the best of our knowledge, such a systematic exhaustive search was not applied to find optimal RMRAs in the past.

*Following are the specific contributions of this paper:*

- An exhaustive search procedure to obtain optimal RMRAs is presented in easy-to-implement algorithmic steps.

- Optimal RMRA configurations for sensor counts from 11 to 15 are reported. As is known, current literature does not have optimal RMRAs beyond $N = 10$. Additionally, near-optimal configurations for sensor counts from 16 to 20 are reported.
- A previously overlooked vulnerability in many 2FRAs is identified: the presence of hidden critical sensors whose failure can disrupt the DCA and compromise the intended purpose of providing two-fold redundancy.

The rest of the paper is organized as follows: Section 2 describes the basic sparse array terminology and the combinatorial optimization problem of finding optimal RMRAs. Section 3 describes the proposed method. Section 4 presents the results and draws comparisons with existing literature. Section 5 discusses the challenges and limitations faced and Section 6concludes the paper with a few future directions.

## 2. Theoretical Background and Problem Formulation

Table 2 outlines the important terminology related to sparse linear arrays, necessary to explain the rest of this paper.

Table 2

Sparse Array Terminology

| Term | Definition |
| --- | --- |
| Physical Array $\mathbb{S}$ | A number set containing the positions of the physical sensors in the array, normalized to half wavelength |
| Aperture $L$ | Separation between the first and the last sensors in the array. |
| Difference Set $\mathbb{Z}$ | The set of all pairwise differences between the elements of $\mathbb{S}$ i.e., $\mathbb{Z} = \{s_i - s_j \mid s_i, s_j \in \mathbb{S}\}$ |
| Difference Coarray (DCA) $\mathbb{D}$ | A set formed with unique elements of $\mathbb{Z}$, sorted in ascending order |
| Spatial Lag $m$ | An entry in the set $\mathbb{D}$ |
| Hole | A missing spatial lag |
| Hole-free DCA | If all spatial lags from $[-L, L]$ are listed in the DCA of an array, it is said to be hole-free |
| Weight $w(m)$ | Number of sensor pairs in $\mathbb{S}$ separated by distance $m$ i.e., number of times a difference $m$ occurs in $\mathbb{Z}$ |
| Doubly Redundant Coarray $\mathbb{D}_2$ | Span of the DCA where all spatial lags have a weight of two or greater |
| Essential Sensor | A physical sensor whose removal from the array affects either the span or continuity of the DCA |
| Fragility ($F$) | Ratio of essential sensors to the total sensors in the array |
| Signal Model | Based on second-order coarray domain processing which is widely discussed in available literature [4], [11], [13] |

By definition, TFRAs must contain only two essential sensors: one positioned at 0 and the other at $L$, thereby yielding a fragility of $2/N$. For instance, in the 7-element RMRA [0, 1, 2, 5, 6, 8, 9], sensors 1, 2, 5, 6, or 8 are non-essential as their individual failure does not impact the span or continuity of the DCA. Contrary to this, the failure of either 0 or 9 reduces the range of the DCA from [-9, 9] to [-8, 8]. Because the sensors at 0 and 9 are essential, the array offers a fragility of 2/7, as expected of RMRAs.

The combinatorial optimization problem (P) of finding optimal RMRAs is defined below

$$\text{(P):} \quad \mathbb{S} \triangleq \text{argmax } L \quad subject\ to$$
$$|\mathbb{S}| = N \qquad |\mathbb{D}| = 2L + 1 \qquad (1)$$
$$|\mathbb{D}_2| = 2L - 1 \quad |\varepsilon_1| = 2 \quad and \quad L \geq N$$

(P) determines the sparse array $\mathbb{S}$ with the largest possible aperture $L$ using $N$ sensors, such that the DCA is continuous from $[-L, L]$ and doubly redundant from $[-(L-1), (L-1)]$, i.e., $|\mathbb{D}_2| = 2L - 1$. The array contains exactly two essential sensors, i.e., $|\varepsilon_1| = 2$. Because a uniform linear array (ULA) with $N$ sensors provides an aperture of $N - 1$, a sparse array must provide an aperture larger than this limit; hence, the constraint $L \geq N$. An array is said to be optimal when a larger array (that satisfies all the properties) cannot be constructed using the available sensors. Section 3 deals with the methods to solve this optimization problem.

## 3. Proposed Method

### 3.1 Demystifying the Optimization Problem

Because the optimization problem outlined in (1) is abstract, it has to be broken down into manageable steps before we could attempt solving it. The first step is to model the array's exact behavior during the healthy and faulty conditions to satisfy the constraint $|\varepsilon_1| = 2$. The weight function could be leveraged to characterize coarray behavior. Specifically, the weight function of the healthy array shall follow the following distribution:

$$\mathbb{S}: w(i) \geq 2; i \in [0, L-1] \tag{2}$$

$$w(L) = 1$$

In a similar way, (3) characterizes the weight function for the sensor failure case

$$\mathbb{S}\setminus\{n\}; n \notin (0, L):$$

$$w(i) \geq 1; i \in [0, L] \tag{3}$$

This modular approach immensely benefits the algorithm designer to develop the problem-solving logic as (2) and (3) clearly delineate the constraints to be satisfied by the array in different scenarios. We propose a special variant of the exhaustive search, namely, leap-on-success exhaustive search to tackle this optimization task.

### 3.2 Pseudocode of the Proposed Approach

A computer program to perform the proposed leap-on-success exhaustive search was developed in Python using the combinations operation from the itertools package. The pseudocode for the same is given below.

---
**Algorithm 1** Exhaustive Search Method to Find Optimal RMRAs
---
**Require:** N      ▷ Desired array size

1: $L \leftarrow N$
---

```
2: while search do
3:     solutionFound ← false    ▷ Flag set to False until solution found
4:     for all A in GETCOMBINATIONS ()   ▷ candidate arrays A that provide aperture L using N sensors
5:         if CONSTRAINTCHECK(A, L, N) then
6:             print "Valid configuration found for L"
7:             solutionFound ← true
8:                 break   ▷ Stop searching for other arrays with this L
9:         end if
10:    end for   ▷ Failed to find L using N sensors
11: L ← L + 1   ▷ Checking for higher L using N sensors
12: end while
```

The function GETCOMBINATIONS uses the `itertools` package in Python and generates all possible positions of *N* sensors that provide an aperture *L*. The candidate arrays are then passed through the CONSTRAINTCHECK function to determine whether they satisfy RMRA properties given in (1). Whenever the first the candidate array that satisfies the RMRA constraints is found, the program exits the current search space and proceeds to the next search space by incrementing *L*, where it explores the search space to find an aperture with an higher aperture with the same number of sensors.

### 3.3 Detailed Working of the Proposed Technique

Considering the NP-hard nature of this problem, we propose a leap-on-success exhaustive search (LoSES) technique, a variant of traditional exhaustive search methods, aimed at achieving optimal solutions with a lower number of computations. Unlike traditional exhaustive search, which evaluates every point in the search space regardless of whether a successful solution is found or not, LoSES halts the search process immediately once a valid solution satisfying all the constraints is found. This preserves the exhaustive nature in principle, while avoiding unnecessary computation when a solution is discovered partway through the search. For instance, if a solution is found after exploring just 10% of the search space, the remaining 90% of the search space could be safely discarded, or vice versa. In any case, it saves the number of computations needed per stage. If a valid array is not found even after exhaustively searching a given solution space, the algorithm reports failure.

An aperture $L$ is obtained when $L + 1$ sensors are placed at grid positions 0, 1, 2, 3,…, $L$, as in a ULA. However, a sparse array does not have sensors at all grid points. Choosing which of the grid points needs to be populated by a sensor forms the essence of finding an optimum sparse array. There are $\binom{L+1}{N}$ choices for this. As sensors are always assigned to grid points 0 and $L$ (for preserving the aperture), only $N - 2$ sensors remain to be assigned at grid points between 0 and $L$. Therefore, a total of $\binom{L-1}{N-2}$ combinations of sensor positions are possible for this aperture.

The search begins by initializing $L = N$. The candidate arrays corresponding to the above combinations are generated and passed to the constraint checking function. If a valid combination is found, the algorithm stops exploring the current search space and increments $L$ to begin a new stage, where it tries to find an array with aperture $L + 1$ using the same number of sensors. With each successful discovery, the search space expands: $[0, L + 1]$, $[0, L + 2]$, and so on. At each stage, the number of combinations grows — $\binom{L}{N-2}$, $\binom{L+1}{N-2}$ etc., increasing the computational effort required. A key limitation is that there is no predefined upper bound on the achievable aperture $L$ for RMRAs, so the algorithm must continue this process until a stage is reached where no valid solution exists. If no solution is found for a given aperture $L + x$, the algorithm returns a failure message, and the last successful solution at $L + x - 1$ is considered optimum. The leap-on-success behavior applies independently at each stage and does not imply early termination of the overall multi-stage process.

## 4. Results

This section explains the numerical results obtained from the program for various array sizes. Additionally, comparisons with other TFRA families are provided. A major drawback in 2FRAs is also highlighted with the help of numerical examples.

*4.1 Validating the Proposed Method*

The smallest $N$ for which a TFRA exists is 6. Although Liu and Vaidyanathan present arrays for $N \geq 4$, the arrays for $N = 4$ and $N = 5$ are not sparse but mere ULAs. Sparse arrays with TFRA properties start from $N = 6$. This is consistent with the principle that a $\beta$-FRA exists for $N \geq 3\beta$ [29]. Hence, we started finding optimum array configurations from $N = 6$.

Table 3 shows the optimal RMRAs for array sizes of six to ten, obtained from our program. The array apertures match the ones proposed by Liu and Vaidyanathan, thereby validating the correctness of our program. It must be noted that multiple valid array configurations might exist for a given pair of $N$ and $L$. The exact positions of sensors do not play a major role as long as the array satisfies the given constraints. Hence, optimality is related to the array aperture and not sensor positions as such.

Table 3

Optimal Arrays for $N = 6$ to $N = 10$

| N | Array Configurations |
|---|---|
| 6 | [0, 1, 2, 3, 5, 6] |
| 7 | [0, 1, 2, 4, 6, 8, 9] |
| 8 | [0, 1, 2, 3, 5, 8, 11, 12] |
| 9 | [0, 1, 2, 3, 4, 9, 10, 14, 15] |
| 10 | [0, 1, 2, 6, 7, 8, 15, 16, 18, 19] |

*4.2 Optimal Arrays for $N = 11$ to $N = 15$*

After establishing the functional correctness of the program, we attempted exploring the optimum RMRA configurations for $N > 10$. Table 4 shows the array configurations returned by the program for $N = 11$.

Table 4

Solutions of (P) for $N = 11$

| Sensor Count | Array Configuration |
|---|---|
|  | [0, 1, 2, 3, 4, 5, 6, 7, 8, 10, 11] |
|  | [0, 1, 2, 3, 4, 5, 6, 7, 8, 11, 12] |
|  | [0, 1, 2, 3, 4, 5, 6, 7, 8, 12, 13] |
|  | [0, 1, 2, 3, 4, 5, 6, 7, 8, 13, 14] |
| $N = 11$ | [0, 1, 2, 3, 4, 5, 6, 7, 8, 14, 15]<br>[0, 1, 2, 3, 4, 5, 6, 7, 8, 15, 16]<br>[0, 1, 2, 3, 4, 5, 6, 8, 10, 16, 17]<br>[0, 1, 2, 3, 4, 5, 6, 8, 11, 17, 18]<br>[0, 1, 2, 3, 4, 5, 6, 11, 12, 18, 19]<br>[0, 1, 2, 3, 4, 5, 6, 12, 13, 19, 20]<br>[0, 1, 2, 3, 4, 5, 6, 13, 14, 20, 21]<br>**[0, 1, 2, 3, 4, 10, 11, 16, 17, 21, 22]** |

The computational behavior of the algorithm can be better understood by considering the case for $N = 11$, in detail. The search starts at $L = 11$. Since sensors are already fixed at grid points 0 and 11, only ten grid points viz., [1, 10] are available to place the remaining nine sensors, leading to $\binom{10}{9} = 10$ possible combinations. If a valid configuration is found, the algorithm quits the current search space and proceeds to the next stage defined by $L = 12$. At this stage, it must choose nine grid positions in the range [1, 11] that satisfy the given constraints. The search proceeds through multiple stages in a similar manner from $L = 11$ to $L = 22$, involving 12 distinct subproblems, each with increasing computational load: $\binom{10}{9}, \binom{11}{9}, \binom{12}{9}, \ldots,$ to $\binom{21}{9}$. At stage 13 ($L = 23$), the algorithm exhaustively checks all $\binom{22}{9} = 497420$ combinations but fails to find a valid solution. As a result, the last successful configuration, found at stage 12 with $L = 22$ (last row of Table 4), is declared as the optimal arrangement for $N = 11$ sensors. Note that this cannot be ascertained unless the program prints: 'Failure to find $L = 23$ for $N = 11$'.

In a similar way, optimum array configurations for $N = 12$ to $N = 15$ have been obtained using the program (see appendix). Of all the arrays obtained, Table 5 presents only the optimum configurations for $N = 11$ to $N = 15$. Table 5 can be considered as the main contribution of this paper because optimal RMRAs beyond $N = 10$ were previously unknown.

Table 5

Optimal RMRAs for array sizes 11 to 15

| N | Array Configurations |
|---|---|
| 11 | [0, 1, 2, 3, 4, 10, 11, 16, 17, 21, 22] |
| 12 | [0, 1, 2, 3, 4, 5, 12, 13, 19, 20, 25, 26] |
| 13 | [0, 1, 2, 4, 5, 9, 14, 19, 24, 25, 30, 31, 32] |
| 14 | [0, 1, 2, 3, 4, 5, 12, 14, 21, 23, 29, 30, 35, 36] |
| 15 | [0, 1, 2, 4, 5, 9, 14, 19, 24, 29, 34, 35, 40, 41, 42] |

## *4.3 Near-Optimal Arrays from $N = 16$ to $N = 20$*

Table 6 lists the RMRA configurations for array sizes 16 to 20. These solutions are classified as near-optimal because the computational resources were exhausted before the entire solution space could be searched. For example, the program could not finish searching the whole solution space for $N = 16$ and $L = 48$. Hence, we cannot conclusively comment whether 47 is the optimum aperture for $N = 16$.

Table 6

Near-Optimal RMRA Configurations

| N | Sensor Positions |
|---|---|
| 16 | [0, 1, 2, 3, 5, 7, 16, 18, 26, 29, 35, 38, 39, 43, 46, 47] |
| 17 | [0, 1, 2, 3, 4, 5, 6, 7, 8, 18, 20, 30, 32, 41, 42, 50, 51] |
| 18 | [0, 1, 2, 3, 4, 5, 6, 7, 8, 9, 20, 22, 33, 35, 45, 46, 55, 56] |
| 19 | [0, 1, 2, 3, 4, 5, 6, 7, 8, 9, 10, 22, 24, 36, 38, 49, 50, 60, 61] |
| 20 | [0, 1, 2, 3, 4, 5, 6, 7, 8, 9, 10, 11, 24, 26, 39, 41, 53, 54, 65, 66] |

Another reason to confirm that the apertures listed in Table 6 are near-optimal is the possibility of being able to obtain higher apertures for $N = 17$ and $N = 19$ through pattern identification. In Fig. 2, the top outermost boxes show the physical sensor positions for $N = 13$ and $N = 15$, as listed in Table 6. From these positions, we derived the inter-element spacing (IES) representation of these two arrays as shown in the top inner circles of Fig. 2. After identifying appropriate patterns in the array structure (repeated spacings of five units), we could extrapolate these to find the IES representations for 17 and 19, as shown in the bottom inner circles of Fig. 2. Finally, the sensor positions in the outermost boxes at the bottom of Fig. 2 were obtained using the extrapolated IES representations. Note that $N = 17$ ends at 52 and $N = 19$ ends at 62, signaling a possibility to attain higher apertures than those found in Table 6, reinforcing the fact that the array configurations listed in Table 6 are indeed near-optimal.

The inability to find optimal configurations for array sizes 16 to 20 can be attributed to limited computational resources and main memory. The results in Table 6 were obtained using a laptop with an Intel i5-1155G7 (2.5 GHz) processor and 16 GB RAM, running the Windows 11 operating system. Nevertheless, if there is access to better computational resources, the program could explore the search space fully and produce optimum solutions.

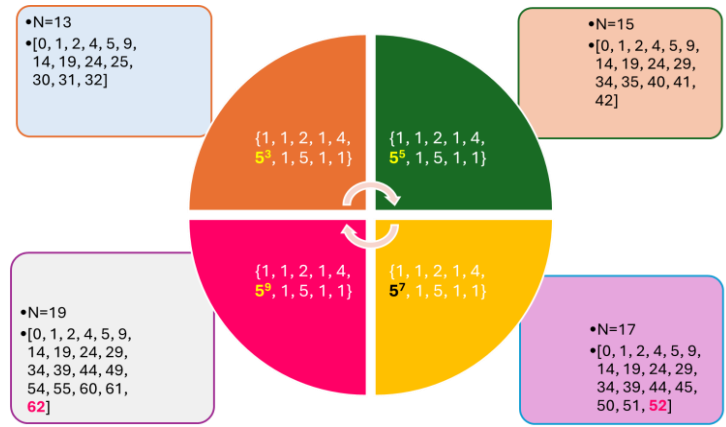

Fig. 2. New Configurations for $N = 17$ and $N = 19$ through IES Notation Analysis and Pattern Finding

### *4.4 Comparison with other known TFRAs*

Table 7 provides a comparison of the apertures offered by TFRAs for sensor counts from six to 20. Specific comparison is against 2FRAs as they have the most successful formulation among all TFRAs.

Table 7

Comparison of TFRA Apertures

| $N$ | SymNA | Aperture ($L$) | |
| --- | --- | --- | --- |
| | | RMRA (Obtained from the proposed program) | 2FRA |
| 6 | ** | 6 | $7_3$ |
| 7 | ** | 9 | $10_4$ |
| 8 | ** | 12 | $13_5$ |
| 9 | ** | 15 | $16_6$ |
| 10 | ** | 19 | $19_{10}$ |
| 11 | ** | 22 | $23_{12}$ |
| 12 | ** | 26 | $27_{14}$ |
| 13 | ** | 32 | $31_{16}$ |
| 14 | ** | 36 | 35 |
| 15 | ** | 42 | 40 |
| 16 | 24 | **47** | 45 |
| 17 | ** | **51** | 50 |
| 18 | 29 | **56** | $55_{28}$ |
| 19 | ** | 61 | $61_{31}$ |
| 20 | 35 | 66 | $67_{34}$ |

**: Array Undefined, $L_s$: Aperture $L$ and critical sensor at $s$

As observed, the symmetric nested array is not defined for most of the array sizes considered here. The 2FRA, though elegant in formulation, suffers from the presence of critical sensors at positions mentioned in the subscript next to each aperture in the rightmost column of Table 7. The failure of a critical sensor causes holes in the DCA, as described in the next subsection. It can be noted that 2FRAs with 14 to 17 sensors do not have critical sensors.

Although near-optimal apertures are reported for array sizes 16 to 18 using the proposed approach, these are higher than the apertures provided by 2FRAs with the same number of sensors. Additionally, the arrays reported are devoid of any critical sensors. Fig. 3 provides a visual representation of the array configurations considered in Table 7 for $N = 16$.

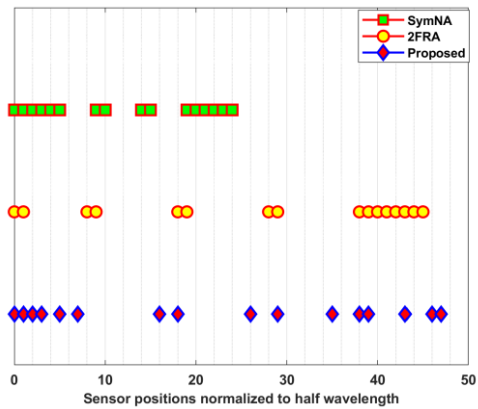

Fig. 3. Sensor positions for various 16-element TFRAs

## *4.5 Problem of Hidden Essential Sensors in 2FRAs*

One might wonder about the potential risks or consequences of having a critical sensor in the array. We illustrate the consequences with a numerical example. Consider the 2FRA configuration for $N = 13$. The array has sensors at [0, 1, 7, 8, 16, 17, 25, 26, 27, 28, 29, 30, 31]. The weight function of this array in the healthy state is shown in Fig. 4a. On the surface, it looks like a TFRA because of the double difference base that it provides. However, detailed failure analysis of the array reveals that the sensor at {16} is critical to the array's functionality as its failure creates a hole in the DCA at spatial lag 15, as shown in Fig. 4b. Nevertheless, individual failure of any other sensor in the above array does not affect the continuity of the DCA.

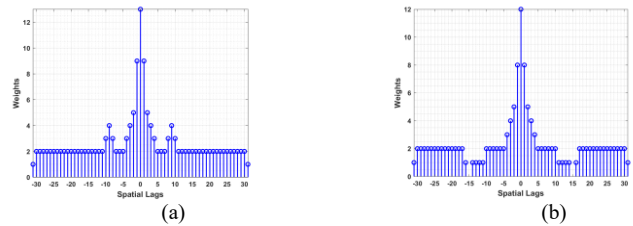

Fig. 4. Weight Function of the 13-element 2FRA in Healthy and Faulty Conditions

Therefore, it can be concluded that this 13-element 2FRA contains a third essential sensor at $16\lambda/2$ in addition to the already agreed ones at 0 and 31. Hence, this array has a fragility of $3/N$ instead of $2/N$. Although the difference in fragilities might seem insignificant, the presence of essential sensors at unagreed locations renders the array unreliable as its robustness is now a function of the faulty sensor's location. While the array can tolerate single sensor failures at almost all locations, the failure of a specific sensor can disrupt its ability to perform accurate DOA estimation, thereby diluting the very purpose of providing two-fold redundancy. This has major implications in various mission critical applications like defense, space exploration, and medical imaging.

## 5. Challenges And Limitations

The primary challenge we encountered was the significant demand for computational resources required by the optimization task, which was expected due to its NP-hard nature. However, what came as a major surprise was the exhaustion of the available main memory capacity. Although strategies such as leap-on-success were employed to accelerate the algorithm, they proved insufficient to manage the increased computational load associated with identifying larger arrays. Even for medium-sized arrays, the search space expanded significantly, both in terms of the number of combinations and the number of stages involved. For instance, in the case of $N = 16$, the final value of $L$ reached as high as 47. Here, the algorithm iterated through 32 stages, from $L = 16$ to $L = 47$, thus exploring 32 distinct subproblems. Due to the exhaustion of computational resources, the algorithm could not evaluate higher apertures such as $L = 48$ and $L = 49$, and thus, we cannot assert that the resulting array is truly optimal. Consequently, only near-optimal apertures are reported for $N = 16$ through $N = 20$.

In the future, the algorithm could be optimized to start at a higher value of $L$. Tighter lower bounds for $L$ could be derived using the fact that optimal TFRAs possess a redundancy $R$ between two and four [29], where $R = \frac{N(N-1)}{2L}$. The upper bound on redundancy, $R < 4$, provides a lower bound on $L$ in terms of the sensor count $N$. Another potential improvement could involve starting the search for $N + 1$ sensors from an $L$-value that was unsuccessful for $N$ sensors, thus reducing the number of stages to be traversed. However, these optimizations are beyond the scope of this work and will be explored in future research.

## 6. Conclusion

This work presents a systematic exhaustive search procedure to determine optimal RMRA configurations. A computer program has been developed for the same and optimal RMRAs for arrays with 11 to 15 sensors have been found successfully. Scalability to higher sensor counts ($N > 15$) was difficult as the search space expanded rapidly with each added sensor, rendering the process computationally intractable. Nevertheless, near-optimal solutions for sensor counts from 16 to 20 have been found. Surprisingly, these near-optimal solutions achieve larger apertures than currently known 2FRAs. This study also uncovered a critical vulnerability in existing 2FRAs: the presence of critical sensors at positions other than 0 and $L$. Failure of such sensors renders the DCA discontinuous, compromising the intended purpose of providing two-fold redundancy. Future work will focus on developing near- or sub-optimal RMRAs based on closed-form expressions for sensor placement, enabling both scalability and mathematical tractability. Simultaneously, formal tools will be developed to verify the presence of additional essential sensors within the array that jeopardize its coarray behavior.

**Appendix**

Valid arrays generated for sensor counts $N = 12$ to $N = 15$ using the proposed leap-on-success exhaustive search (LoSES) are listed in Table A.1 below.

Table A.1

Valid TFRA/RMRA Configurations for Array Sizes 12 to 15

| N | Array Configurations |
|---|---|
| 12 | [0, 1, 2, 3, 4, 5, 6, 7, 8, 9, 11, 12] |
| | [0, 1, 2, 3, 4, 5, 6, 7, 8, 9, 12, 13] |
| | [0, 1, 2, 3, 4, 5, 6, 7, 8, 9, 13, 14] |
| | [0, 1, 2, 3, 4, 5, 6, 7, 8, 9, 14, 15] |
| | [0, 1, 2, 3, 4, 5, 6, 7, 8, 9, 15, 16] |
| | [0, 1, 2, 3, 4, 5, 6, 7, 8, 9, 16, 17] |
| | [0, 1, 2, 3, 4, 5, 6, 7, 8, 9, 17, 18] |
| | [0, 1, 2, 3, 4, 5, 6, 7, 9, 11, 18, 19] |
| | [0, 1, 2, 3, 4, 5, 6, 7, 9, 12, 19, 20] |
| | [0, 1, 2, 3, 4, 5, 6, 7, 12, 13, 20, 21] |
| | [0, 1, 2, 3, 4, 5, 6, 7, 13, 14, 21, 22] |
| | [0, 1, 2, 3, 4, 5, 6, 7, 14, 15, 22, 23] |
| | [0, 1, 2, 3, 4, 5, 6, 7, 15, 16, 23, 24] |
| | [0, 1, 2, 3, 4, 5, 12, 13, 18, 19, 24, 25] |
| | **[0, 1, 2, 3, 4, 5, 12, 13, 19, 20, 25, 26]** |
| 13 | [0, 1, 2, 3, 4, 5, 6, 7, 8, 9, 10, 12, 13] |
| | [0, 1, 2, 3, 4, 5, 6, 7, 8, 9, 10, 13, 14] |
| | [0, 1, 2, 3, 4, 5, 6, 7, 8, 9, 10, 14, 15] |
| | [0, 1, 2, 3, 4, 5, 6, 7, 8, 9, 10, 15, 16] |
| | [0, 1, 2, 3, 4, 5, 6, 7, 8, 9, 10, 16, 17] |
| | [0, 1, 2, 3, 4, 5, 6, 7, 8, 9, 10, 17, 18] |
| | [0, 1, 2, 3, 4, 5, 6, 7, 8, 9, 10, 18, 19] |
| | [0, 1, 2, 3, 4, 5, 6, 7, 8, 9, 10, 19, 20] |
| | [0, 1, 2, 3, 4, 5, 6, 7, 8, 10, 12, 20, 21] |
| | [0, 1, 2, 3, 4, 5, 6, 7, 8, 10, 13, 21, 22] |
| | [0, 1, 2, 3, 4, 5, 6, 7, 8, 13, 14, 22, 23] |
| | [0, 1, 2, 3, 4, 5, 6, 7, 8, 14, 15, 23, 24] |
| | [0, 1, 2, 3, 4, 5, 6, 7, 8, 15, 16, 24, 25] |
| | [0, 1, 2, 3, 4, 5, 6, 7, 8, 16, 17, 25, 26] |
| | [0, 1, 2, 3, 4, 5, 6, 7, 8, 17, 18, 26, 27] |
| | [0, 1, 2, 3, 4, 5, 6, 13, 14, 20, 21, 27, 28] |
| | [0, 1, 2, 3, 4, 5, 6, 14, 15, 21, 22, 28, 29] |
| | [0, 1, 2, 3, 4, 5, 6, 14, 15, 22, 23, 29, 30] |
| | [0, 1, 2, 3, 4, 10, 12, 18, 20, 25, 26, 30, 31] |
| | **[0, 1, 2, 4, 5, 9, 14, 19, 24, 25, 30, 31, 32]** |
| 14 | [0, 1, 2, 3, 4, 5, 6, 7, 8, 9, 10, 11, 13, 14] |
| | [0, 1, 2, 3, 4, 5, 6, 7, 8, 9, 10, 11, 14, 15] |
| | [0, 1, 2, 3, 4, 5, 6, 7, 8, 9, 10, 11, 15, 16] |
| | [0, 1, 2, 3, 4, 5, 6, 7, 8, 9, 10, 11, 16, 17] |
| | [0, 1, 2, 3, 4, 5, 6, 7, 8, 9, 10, 11, 17, 18] |
| | [0, 1, 2, 3, 4, 5, 6, 7, 8, 9, 10, 11, 18, 19] |
| | [0, 1, 2, 3, 4, 5, 6, 7, 8, 9, 10, 11, 19, 20] |
| | [0, 1, 2, 3, 4, 5, 6, 7, 8, 9, 10, 11, 20, 21] |
| | [0, 1, 2, 3, 4, 5, 6, 7, 8, 9, 10, 11, 21, 22] |
| | [0, 1, 2, 3, 4, 5, 6, 7, 8, 9, 11, 13, 22, 23] |
| | [0, 1, 2, 3, 4, 5, 6, 7, 8, 9, 11, 14, 23, 24] |
| | [0, 1, 2, 3, 4, 5, 6, 7, 8, 9, 14, 15, 24, 25] |
| | [0, 1, 2, 3, 4, 5, 6, 7, 8, 9, 15, 16, 25, 26] |
| | [0, 1, 2, 3, 4, 5, 6, 7, 8, 9, 16, 17, 26, 27] |
| | [0, 1, 2, 3, 4, 5, 6, 7, 8, 9, 17, 18, 27, 28] |
| | [0, 1, 2, 3, 4, 5, 6, 7, 8, 9, 18, 19, 28, 29] |
| | [0, 1, 2, 3, 4, 5, 6, 7, 8, 9, 19, 20, 29, 30] |
| | [0, 1, 2, 3, 4, 5, 6, 7, 14, 15, 22, 23, 30, 31] |
| | [0, 1, 2, 3, 4, 5, 6, 7, 15, 16, 23, 24, 31, 32] |
| | [0, 1, 2, 3, 4, 5, 6, 7, 16, 17, 24, 25, 32, 33] |
| | [0, 1, 2, 3, 4, 5, 6, 7, 16, 17, 25, 26, 33, 34] |
| | [0, 1, 2, 3, 4, 5, 12, 13, 20, 21, 28, 29, 34, 35] |
| | **[0, 1, 2, 3, 4, 5, 12, 14, 21, 23, 29, 30, 35, 36]** |
| 15 | [0, 1, 2, 3, 4, 5, 6, 7, 8, 9, 10, 11, 12, 14, 15] |
| | [0, 1, 2, 3, 4, 5, 6, 7, 8, 9, 10, 11, 12, 15, 16] |
| | [0, 1, 2, 3, 4, 5, 6, 7, 8, 9, 10, 11, 12, 16, 17] |
| | [0, 1, 2, 3, 4, 5, 6, 7, 8, 9, 10, 11, 12, 17, 18] |
| | [0, 1, 2, 3, 4, 5, 6, 7, 8, 9, 10, 11, 12, 18, 19] |
| | [0, 1, 2, 3, 4, 5, 6, 7, 8, 9, 10, 11, 12, 19, 20] |
| | [0, 1, 2, 3, 4, 5, 6, 7, 8, 9, 10, 11, 12, 20, 21] |
| | [0, 1, 2, 3, 4, 5, 6, 7, 8, 9, 10, 11, 12, 21, 22] |
| | [0, 1, 2, 3, 4, 5, 6, 7, 8, 9, 10, 11, 12, 22, 23] |
| | [0, 1, 2, 3, 4, 5, 6, 7, 8, 9, 10, 11, 12, 23, 24] |
| | [0, 1, 2, 3, 4, 5, 6, 7, 8, 9, 10, 12, 14, 24, 25] |
| | [0, 1, 2, 3, 4, 5, 6, 7, 8, 9, 10, 12, 15, 25, 26] |
| | [0, 1, 2, 3, 4, 5, 6, 7, 8, 9, 10, 15, 16, 26, 27] |
| | [0, 1, 2, 3, 4, 5, 6, 7, 8, 9, 10, 16, 17, 27, 28] |
| | [0, 1, 2, 3, 4, 5, 6, 7, 8, 9, 10, 17, 18, 28, 29] |
| | [0, 1, 2, 3, 4, 5, 6, 7, 8, 9, 10, 18, 19, 29, 30] |
| | [0, 1, 2, 3, 4, 5, 6, 7, 8, 9, 10, 19, 20, 30, 31] |
| | [0, 1, 2, 3, 4, 5, 6, 7, 8, 9, 10, 20, 21, 31, 32] |
| | [0, 1, 2, 3, 4, 5, 6, 7, 8, 9, 10, 21, 22, 32, 33] |

| |
|---|
| [0, 1, 2, 3, 4, 5, 6, 7, 8, 14, 16, 24, 29, 33, 34] |
| [0, 1, 2, 3, 4, 5, 6, 7, 8, 16, 17, 25, 26, 34, 35] |
| [0, 1, 2, 3, 4, 5, 6, 7, 8, 17, 18, 26, 27, 35, 36] |
| [0, 1, 2, 3, 4, 5, 6, 7, 8, 18, 19, 27, 28, 36, 37] |
| [0, 1, 2, 3, 4, 5, 6, 7, 8, 18, 19, 28, 29, 37, 38] |
| [0, 1, 2, 3, 4, 5, 6, 13, 15, 22, 24, 31, 32, 38, 39] |
| [0, 1, 2, 3, 4, 5, 6, 14, 15, 23, 24, 32, 33, 39, 40] |
| [0, 1, 2, 3, 4, 5, 6, 14, 16, 24, 26, 33, 34, 40, 41] |
| **[0, 1, 2, 4, 5, 9, 14, 19, 24, 29, 34, 35, 40, 41, 42]** |